# Phase Operator Problem
# and Macroscopic Extension of Quantum Mechanics

MASANAO OZAWA

*School of Informatics and Sciences, Nagoya University, Nagoya 464-01, Japan*

## Abstract

To find the Hermitian phase operator of a single-mode electromagnetic field in quantum mechanics, the Schrödinger representation is extended to a larger Hilbert space augmented by states with infinite excitation by nonstandard analysis. The Hermitian phase operator is shown to exist on the extended Hilbert space. This operator is naturally considered as the controversial limit of the approximate phase operators on finite dimensional spaces proposed by Pegg and Barnett. The spectral measure of this operator is a Naimark extension of the optimal probability operator-valued measure for the phase parameter found by Helstrom. Eventually, the two promising approaches to the statistics of the phase in quantum mechanics is synthesized by means of the Hermitian phase operator in the macroscopic extension of the Schrödinger representation.

## 1.  Introduction

The existence and properties of a Hermitian operator on a Hilbert space corresponding to the phase of the electromagnetic field has provoked many discussions since Dirac [1] first discussed the problem. According to the uniqueness theorem of the irreducible representations of the canonical commutation relation due to von Neumann, the commutation relation between the number operator and the phase operator which Dirac presupposed from the cor-respondence between the commutator and the classical Poisson brackets cannot be realized. Further, Susskind and Glogower [2] clearly demonstrated that the polar decomposition of the annihilation operator into the unitary operator of the exponential of the phase and the square root of the number operator presupposed by Dirac is also impossible. However, Pegg and Barnett [3]–[5] recently made an interesting proposal for the problem. They constructed approximate Hermitian phase operators on finite dimensional spaces and claimed that the statistics obtained by their operator approaches the statistics of the phase as the dimension tends to infinity. However, they have failed to find the Hermitian phase operator on an infinite dimensional space as the limit of their approximate operators.

On the other hand, another approach to the problem has been established in quantum estimation theory [6, 7]. This theory discusses optimization problems of quantum measurements quite generally. The statistics of measurement is represented in this theory by a probability operator-valued measure (POM) on a Hilbert space which extends the conventional description by a Hermitian operator. In this approach, the optimum POM of the estimation problem of the phase parameter was found by Helstrom [8], and mathematically rigorous development of this approach is given by Holevo [7].

A promising aspect shared by these two approaches is that the statistics of the phase obtained by the limit process of Pegg and Barnett coincides with the one represented by the optimum POM of the phase parameter [9]–[12]. This shows, however, that contrary to their claim the limit of the exponentials of the approximate phase operators is nothing but the well-known Susskind-Glogower exponential phase operator [2], as long as the limit is taken on the Hilbert space of quantum states with the weak operator topology. The limit in the weak operator topology does not preserve the product operation and hence demolishes the desired properties of the limit operator which Pegg and Barnett [3] described intuitively. According to the Naimark theorem, every POM can be extended to a projection-valued measure on a larger Hilbert space which gives rise to a Hermitian operator by the spectral theory representing an observable in the standard formulation of quantum mechanics. This suggests that there exists the Hermitian phase operator somewhere beyond the Hilbert space of quantum states. Thus in order to realize the intuitive limit of the approximate phase operators, we need an alternative mathematical construction other than the limit on a Hilbert space.

In this paper the attempt from nonstandard analysis outlined in [12] is developed for this purpose. The nonstandard analysis was invented by Robinson [13] and has yielded rigorous and fruitful mathematics of infinite and infinitesimal numbers. We construct a natural extension of the Schrödinger representation and show that the desired Hermitian phase operator exists on this extended Hilbert space. The Hilbert space of this extension of the Schrödinger representation is the direct sum of the original space of quantum states and the space of states with infinite excitation which are naturally considered as the classical limits of the ordinary quantum states.

In the conventional approach, microscopic properties and macroscopic properties are discussed separately in quantum mechanics and in classical mechanics. Although the correspondence principle bridges both mechanics by the mathematical process of taking the limit, this approach cannot describe the quantum mechanical coherence between microscopic states and macroscopic states. Our new representation realizes such a coherent description of quantum and classical mechanics; it is in such a representation that the phase operator behaves as a Hermitian operator. Obviously, the present method is applicable to other difficulties in quantum mechanics concerning the bounded dimensionless quantities such as the rotation angle and the quantities associated with them such as the time of periodic motions. A potential application of this representation other than the above problems is the measurement problem, where the unitary time evolution in an amplifier evolves from a quantum state to a state with infinite excitation [14]. Applications to these problems will be discussed in the forthcoming papers.

For bibliography on the phase operator problem we shall refer to the references of [4, 15, 16], and for the recent developments emerged by the Pegg-Barnett proposal the references of [17]. For quantum estimation theory [6, 7], for quantum measurement theory [18, 19], and for operator algebras [20]. For basic methods of nonstandard analysis, we shall refer to [21]. Applications of nonstandard analysis to physics is not new and has been developed in such papers as [22]–[34], and in monograph [35].

## 2. The Susskind-Glogower operators

The single-mode electromagnetic field is a well-known physical system which is modeled by the quantum mechanical harmonic oscillator with unit mass. Let $\mathcal{H}$ be the Hilbert space of the Schrödinger representation of the quantum mechanical harmonic oscillator. Let $\hat{q}$ and $\hat{p}$ be the position and momentum operators on $\mathcal{H}$. The annihilation operator $\hat{a}$ is defined by

$$\hat{a} = \frac{1}{\sqrt{2\hbar\omega}}(\omega\hat{q} + i\hat{p}), \tag{2.1}$$

where $\omega$ is the angular frequency, and its adjoint $\hat{a}^\dagger$ is the creation operator. Then the number operator $\hat{N}$ is defined by

$$\hat{N} = \hat{a}^\dagger\hat{a}. \tag{2.2}$$

The number operator $\hat{N}$ has the complete orthonormal basis $\{|n\rangle \mid n = 1, 2, \ldots\}$ of $\mathcal{H}$ for which $\hat{N}|n\rangle = n|n\rangle$. The Hamiltonian $\hat{H}$ of the system is given by $\hat{H} = \hbar\omega(\hat{N} + \frac{1}{2})$.

In his original description of the quantized electromagnetic field, Dirac [1] postulated the existence of a Hermitian phase operator $\hat{\phi}_D$ such that the unitary exponential operator $\exp i\hat{\phi}_D$ of $\hat{\phi}_D$ would appear in the polar decomposition of the annihilation operator

$$\hat{a} = (\exp i\hat{\phi}_D)\hat{N}^{-1/2}. \tag{2.3}$$

The difficulty with this approach were clearly pointed out by Susskind and Glogower [2] by showing that the polar decomposition of $\hat{a}$ can be realized by no unitary operators. Instead, they introduced the partial isometries representing the exponentials of the phase

$$\widehat{\exp}_{SG} i\phi = (\hat{N}+1)^{-1/2}\hat{a}, \tag{2.4}$$

$$\widehat{\exp}_{SG} -i\phi = \hat{a}^\dagger(\hat{N}+1)^{-1/2}, \tag{2.5}$$

and the Hermitian operators representing the sine and cosine of the phase

$$\widehat{\cos}_{SG}\phi = \frac{1}{2}(\widehat{\exp}_{SG} i\phi + \widehat{\exp}_{SG} -i\phi), \tag{2.6}$$

$$\widehat{\sin}_{SG}\phi = \frac{1}{2i}(\widehat{\exp}_{SG} i\phi - \widehat{\exp}_{SG} -i\phi). \tag{2.7}$$

Note that the places of the carets in these Susskind-Glogower (SG) operators suggest that these operators are *not* derived by the function calculus of a certain Hermitian operator corresponding to $\phi$. These operators are considered to behave well in the classical limit, but they fail to define well-behaved operators even for periodic functions of the phase in the quantum regime. Thus we cannot derive the correct statistics of the phase from these operators. However, it turns out that these operators give the correct mean values of the corresponding quantities $e^{\pm i\phi}$, $\sin\phi$ and $\cos\phi$. A systematic method for obtaining the correct statistics needs a new mathematical concept generalizing the Hermitian operators, which is described in the next section.

## 3. Quantum estimation theory

Denote by $\mathcal{L}(\mathcal{H})$ the algebra of bounded operators on $\mathcal{H}$. In the conventional approach [36], any observable $A$ has a unique Hermitian operator $\hat{A}$ on $\mathcal{H}$, and any Hermitian operator $\hat{A}$ has a unique resolution of the identity $E_{\hat{A}}(x)$ ($x \in \mathbf{R}$), called the *spectral family* of $\hat{A}$. Then the probability distribution function of $A$ in a (normalized) state $\psi \in \mathcal{H}$ is given by

$$\Pr[A \le x|\psi] = \langle\psi|E_{\hat{A}}(x)|\psi\rangle. \tag{3.1}$$

For any Borel function $f$, the observable $f(A)$ has the Hermitian operator $f(\hat{A})$ defined by the function calculus

$$f(\hat{A}) = \int_{\mathbf{R}} f(x) \, dE_{\hat{A}}(x), \tag{3.2}$$

and then the mean value of $f(A)$ in state $\psi$ is

$$\mathrm{Ex}[f(A)|\psi] = \langle \psi | f(\hat{A}) | \psi \rangle. \tag{3.3}$$

This framework of the statistical interpretation based on the Hermitian operator can be generalized to the framework based on the so-called POM, by now fairly well-known, as follows.

A *POM* or *non-orthogonal resolution of the identity* on $\mathcal{H}$ is a family $P(x)$ $(x \in \mathbf{R})$ of bounded Hermitian operators on $\mathcal{H}$ with the following conditions:

(S1) $\lim_{x \to -\infty} P(x) = 0$, $\lim_{x \to \infty} P(x) = 1$, and $P(x) = \lim_{n \to \infty} P(x + n^{-1})$, where the limit is taken in the strong operator topology.

(S2) From $x' < x''$ it follows that $0 \leq P(x) \leq P(x') \leq 1$.

Note that it is usual that "POM" abbreviates the "probability operator-valued measure", which is more or less mathematically equivalent to the simpler notion of "non-orthogonal resolution of the identity", as long as we restrict our attention to one-dimensional probability distributions.

Now, we shall start with the following presupposition: *corresponding to any measurable physical quantity $X$, there is a unique POM $P_X$ such that the probability distribution function of $X$ in a (normalized) state $\psi \in \mathcal{H}$ is given by*

$$\Pr[X \leq x | \psi] = \langle \psi | P_X(x) | \psi \rangle. \tag{3.4}$$

For an observable $A$ in the conventional framework, the corresponding POM $P_A$ is the spectral resolution $E_{\hat{A}}$ of the Hermitian operator $\hat{A}$. For any Borel function $f$, the operator $\widehat{f(X)} = \int f \, dP_X$ is defined as follows. Let $\mathrm{dom}(\widehat{f(X)})$ be the set such that

$$\mathrm{dom}(\widehat{f(X)}) = \{ \eta \in \mathcal{H} | \int_{\mathbf{R}} |f(x)|^2 \, d\langle \eta | P_X(x) | \eta \rangle < \infty \}.$$

For any $\eta \in \mathrm{dom}(\widehat{f(X)})$, $\widehat{f(X)}\eta$ is defined by the relation

$$\langle \xi | \widehat{f(X)} \eta \rangle = \int_{\mathbf{R}} f(x) d \langle \xi | P_X(x) | \eta \rangle,$$

for all $xi \in \mathcal{H}$, where the integral is Lebesgue-Stieltjes integral. The mean value of the quantity $f(X)$ in state $\psi \in \mathrm{dom}(\widehat{f(X)})$ is given by

$$\mathrm{Ex}[f(X)|\psi] = \langle \psi | \widehat{f(X)} | \psi \rangle. \tag{3.5}$$

Then this generalizes the function calculus (3.2) based on a spectral resolution. Namely, $P_X$ is an orthogonal resolution, i.e., $P_X(x)$ is a projection for each $x \in \mathbf{R}$, if and only if $\hat{X} = \int x \, dP_X$ is a Hermitian operator on $\mathcal{H}$ such that $E_{\hat{X}} = P_X$ and $f(\hat{X}) = \widehat{f(X)}$ for any Borel function $f$.

By a measurement of quantity $X$, we mean any experiment the outcome of which is predicted by the probability distribution function given by (3.4). For measurability of quantity $X$ with POM $P_X(x)$, it is known that for any POM $P_X(x)$ on $\mathcal{H}$, there is another Hilbert space $\mathcal{K}$, a unit vector $\xi \in \mathcal{K}$, a unitary operator $U$ on $\mathcal{H} \otimes \mathcal{K}$, and a Hermitian operator $\hat{A}$ on $\mathcal{K}$ with spectral resolution $E_{\hat{A}}$ satisfying

$$\langle \psi | P_X(x) | \psi \rangle = \langle \psi \otimes \xi | U^{\dagger}(1 \otimes E_{\hat{A}}(x)) U | \psi \otimes \xi \rangle, \tag{3.6}$$

for all $\psi \in \mathcal{H}$ [37]. This experiment consists of the following process; 1) preparation of the apparatus (described by $\mathcal{K}$) in state $\xi$, 2) interaction (described by $U$) between the object (described by $\mathcal{H}$) and the apparatus, 3) measurement of the observable $A$ (corresponding to $E_A(x)$) in the apparatus. We shall call any experiment with the above process which satisfies (3.6) as a *measurement* of quantity $X$ or POM $P_X$. Thus for any quantity $X$ with POM $P_X$ there is a measurement the outcome of which is predicted by (3.4). Note that this interpretation of the statistics of the outcome of the measurement is a natural consequence of the conventional postulate (3.1). Our presupposition is thus a conservative extension of the conventional formulation of quantum mechanics, in the sense that, if every observable with the Hermitian operator can be measured, so can every quantity with the POM.

The determination of the statistics of the phase is thus reduced to the determination of the POM $P_\phi$ corresponding to the phase $\phi$. This problem is solved in quantum estimation theory as follows. We assume that the phase $\phi$ has values in $[0, 2\pi)$, and hence we require $P_\phi(x) = 0$ for $x < 0$ and $P_\phi(2\pi) = 1$. Since the phase is canonically conjugate to the action in classical mechanics, the number operator is the infinitesimal generator of the phase shift operators $e^{i\theta \hat{N}}$ in quantum mechanics. Thus the POM $P_\phi$ should satisfy the relations

$$e^{-i\theta \hat{N}} dP_\phi(\theta') e^{i\theta \hat{N}} = dP_\phi(\theta' \ominus \theta),$$

$$0 \leq \theta', \ \theta, \theta' \ominus \theta < 2\pi, \quad \theta' \ominus \theta \equiv \theta' - \theta \pmod{2\pi}.$$

(3.7)

Or equivalently,

$$e^{-i\theta \hat{N}} \bar{P}_\phi(B) e^{i\theta \hat{N}} = \bar{P}_\phi(B_{-\theta}),$$

$$B, B_{-\theta} \subset [0, 2\pi), \quad B_{-\theta} = B - \theta \pmod{2\pi},$$

(3.8)

where $\bar{P}_\phi(B) = \int_B dP_\phi$. Any POM satisfying (3.7) is called a *covariant* POM [7]. It is well known that there is no Hermitian operator such that its spectral resolution satisfies the above relations, but there are many solutions among general POM's. In order to select the optimum one, consider the following estimation problem of the phase parameter $\theta$. Let us given an optical mode in a reference state $\psi \in \mathcal{H}$ which is supposed to interact with a phase shifter with unknown shift parameter $\theta$ ($0 \leq \theta < 2\pi$) so that the outgoing state is $\psi_\theta = e^{i\theta \hat{N}} \psi$. The estimation problem is to find an experiment in state $\psi_\theta$ which gives the best estimate of the parameter $\theta$. This is equivalent to find a measurement in the state $\psi_\theta$ the outcome $\bar{\theta}$ of which is the best estimate of the parameter $\theta$. The relevance of this estimation problem to the determination of the POM for the phase is as follows. Suppose that the reference state $\psi$ were the phase eigenstate $\psi = |\phi = 0\rangle$. The outgoing state from the phase shifter would be the phase eigenstate $\psi_\theta = |\phi = \theta\rangle$, for which the the best estimator would give the estimate $\bar\theta = \theta$ with probability 1. Thus in this case the best estimate results from the *measurement of the phase*. Thus for a POM $P$ to represent the phase, it is necessary that it is the optimum estimator of this estimation problem if the reference state approximates the phase eigenstate closely. Whereas we do not know what and where are phase eigenstates, we can reach the *essentially unique* solution as follows. For a POM $P$, the joint probability distribution

$$p(d\theta, d\bar\theta) = \langle \psi_\theta | dP(\bar\theta) | \psi_\theta \rangle \, \frac{d\theta}{2\pi} \tag{3.9}$$

gives naturally the joint probability distribution of the true parameter $\theta$ and the estimate $\bar\theta$. Given an appropriate error function $W(\theta - \bar\theta)$, which gives the penalty for the case $\theta \neq \bar\theta$, the optimum estimator should minimize the average error

$$\int_0^{2\pi} W(\theta - \bar\theta) \, p(d\theta, d\bar\theta). \tag{3.10}$$

Optimization problems of this type have been studied extensively in quantum estimation theory [6, 7]. The following POM $P_{opt}$ is the covariant POM which is the common optimum solution for a large class of error functions such as $W(x) = 4 \sin^2 \frac{x}{2}$ or $W(x) = -\delta(x)$, where $\delta(x)$ is the periodic $\delta$-function:

$$\langle n | dP_{opt}(\theta) | n' \rangle = e^{i(\alpha_n - \alpha_{n'})} e^{i(n-n')\theta} \, \frac{d\theta}{2\pi}, \tag{3.11}$$

where $|n\rangle$ $(n = 0, 1, \ldots)$ is the number basis and $\alpha_n = \arg\langle n | \psi \rangle$. Note that an optimum solution for the particular error function $-\delta(x)$ is called a *maximum likelihood* estimator, and for an arbitrary error function $W$ a *Bayes* estimator for $W$. Since this problem is not of the estimation of the absolute phase of the outgoing state, the optimum solutions depend on the phase factors $\alpha_n$ of the reference state $\psi$. However, this dependence only reflects the our optional choice of the phase eigenstate $|\phi = 0\rangle$, and each choice of the optimum

POM $P_{opt}$ determines a unique $|\phi = 0\rangle$ among *physically equivalent* alternatives. To see this physical equivalence, replace each number state $|n\rangle$ by the physically equivalent $e^{i\alpha_n}|n\rangle$, and the same POM $P_{opt}$ turns to be the solution for $\alpha_n = 1$. Thus the particular choice of $\alpha_n$ does not affect the physics. For simplicity, we choose the solution for $\alpha_n = 1$ ($n = 0, 1, \ldots$) and determine it as the POM $P_\phi$ for the phase $\phi$, i.e.,

$$\langle n|dP_\phi(\theta)|n'\rangle = e^{i(n-n')\theta}\,\frac{d\theta}{2\pi}. \tag{3.12}$$

We call $P_\phi$ as the *phase POM*.

Applying (3.4) and (3.5) to the phase POM, we obtain the statistics of the phase $\phi$ as follows. The probability distribution function of the phase $\phi$ in state $\psi = \sum_n c_n|n\rangle \in \mathcal{H}$ is

$$\Pr[\phi \le \theta|\psi] = \sum_{n,n'} c_n c_{n'}^* \int_0^\theta e^{i(n-n')\theta}\,\frac{d\theta}{2\pi}.$$

The mean value of the Borel function $f(\phi)$ of the phase $\phi$ is $\langle\psi|\widehat{f(\phi)}|\psi\rangle$, where we have

$$\begin{aligned}
\widehat{f(\phi)} &= \int_0^{2\pi} f(\theta)dP_\phi(\theta) \\
&= \sum_{n,n'} |n\rangle \int_0^{2\pi} e^{i(n-n')\theta} f(\theta)\,\frac{d\theta}{2\pi}\langle n'|.
\end{aligned}$$

An interesting result from this is that the SG operators coincide with the operators defined from the phase POM [7, p. 141], i.e.,

$$\widehat{\exp}_{SG}\pm i\phi = \int_0^{2\pi} e^{\pm i\theta}\,dP_\phi(\theta) = \widehat{e^{\pm i\phi}}, \tag{3.13}$$

$$\widehat{\cos}_{SG}\phi = \int_0^{2\pi} \cos\theta\,dP_\phi(\theta) = \widehat{\cos}\phi, \tag{3.14}$$

$$\widehat{\sin}_{SG}\phi = \int_0^{2\pi} \sin\theta\,dP_\phi(\theta) = \widehat{\sin}\phi. \tag{3.15}$$

Thus the SG operators give the correct mean values of $f(\phi) = e^{\pm i\phi}$, $\sin\phi$ and $\cos\phi$, but none of their powers.

The phase POM $P_\phi$ gives the correct mean value for all Borel functions of the phase, but it gives little information about the algebraic structures of the physical quantities including the phase.

## 4.   The Pegg-Barnett operators

Now we shall turn to the proposal due to Pegg and Barnett [3]. They start with the $s$-dimensional subspace $\Psi_s$ of $\mathcal{H}$ spanned by number states $|n\rangle$ with $n = 0, 1, \ldots, s-1$. For $\theta_m = m\Delta\theta$ $(m = 0, 1, \ldots, s-1)$, where $\Delta\theta = 2\pi/s$, the *approximate phase state* is

$$|\theta_m\rangle = s^{-1/2} \sum_{n=0}^{s-1} e^{in\theta_m}|n\rangle, \tag{4.1}$$

and the *approximate phase operator* $\hat{\phi}_s$ on $\Psi_s$ is

$$\hat{\phi}_s = \sum_{m=0}^{s-1} \theta_m |\theta_m\rangle\langle\theta_m|. \tag{4.2}$$

Their intrinsic proposal is that the mean value of the quantity $f(\phi)$ in state $\psi$ is the limit of $\langle\psi|f(\hat{\phi}_s)|\psi\rangle$ as $s \to \infty$. Then for state $\psi = \sum_{n=0}^{k} c_n |n\rangle$ $(k < \infty)$, we have

$$
\begin{aligned}
\lim_{s\to\infty} \langle\psi|f(\hat{\phi}_s)|\psi\rangle &= \sum_{n,n'=0}^{k} c_n c_{n'}^* \lim_{s\to\infty} \sum_{m=0}^{s-1} f(\theta_m) e^{i(n-n')\theta_m} \frac{\Delta\theta}{2\pi} \\
&= \sum_{n,n'=0}^{k} c_n c_{n'}^* \int_0^{2\pi} f(\theta) e^{i(n-n')\theta} \frac{d\theta}{2\pi} \\
&= \int_0^{2\pi} f(\theta) \langle\psi|P_\phi(d\theta)|\psi\rangle \\
&= \langle\psi|\widehat{f(\phi)}|\psi\rangle,
\end{aligned}
$$

for any continuous function $f(\theta)$ on $[0, 2\pi]$. Thus, the mean values are the same as those given by the phase POM $P_\phi(d\theta)$, and we have

$$\lim_{s\to\infty} f(\hat{\phi}_s) = \int_0^{2\pi} f(\theta) P_\phi(d\theta) = \widehat{f(\phi)}, \tag{4.3}$$

where the limit is taken in the weak operator topology. In particular, the limit of their exponential, sine and cosine phase operators are the SG operators, i.e.,

$$
\begin{aligned}
\lim_{s\to\infty} \exp\pm i\hat{\phi}_s &= \widehat{\exp}_{SG}\pm i\phi, \\
\lim_{s\to\infty} \cos\hat{\phi}_s &= \widehat{\cos}_{SG}\phi, \\
\lim_{s\to\infty} \sin\hat{\phi}_s &= \widehat{\sin}_{SG}\phi.
\end{aligned}
$$

Therefore, the statistics of the phase obtained by Pegg and Barnett coincides with the statistics obtained by the phase POM, and that the limit of exponentials of the approximate phase operators on finite dimensional spaces is nothing but the SG exponential operators, as long as the limit is taken in the weak operator topology on the Hilbert space $\mathcal{H}$.

In the following sections, we shall develop an entirely new approach to the limit of the approximate phase operators.

## 5. Nonstandard analysis

It seems that certain amount of a physical quantity can be described both in quantum mechanics and classical mechanics consistently. Consider a highly excited single-mode radiation field with the number $n$ of quanta. Let $E_q$ the energy of this radiation field in quantum mechanics with unit $U_q$, and $E_c$ in classical mechanics with unit $U_c$. Then, if both mechanics describe the same physical state, we should put $E_q U_q = E_c U_c$. According to quantum mechanics, we have $E_q = \hbar_q \omega_q (n_q + \frac{1}{2})$. We can choose the unit system of quantum mechanics so that $\hbar_q = 1$, and the unit of the time is assumed to be common in both mechanics, so that $\omega_q = \omega_c$. In classical mechanics, the position-momentum uncertainty is negligibly small, and hence it is necessary to choose the unit system of classical mechanics such that $\hbar_c$ is negligibly small. In classical mechanics, the relation $E_c = \hbar_c \omega_c (n_c + \frac{1}{2})$ is consistent with the fact that the classical energy $E_c$ is a continuous variable, since the minimum increment $\hbar_c \omega_c$ is negligibly small. However, in order for $E_c$ to be a finite positive number, $n_c$ should be infinitely large. If we do not invoke negligibly small $\hbar_c$ nor infinitely large $n_c$ in classical mechanics, nothing appears to be inconsistent. However, a question arise—can we assume that $n_q$ is finite and yet $n_c$ is infinite? Since the number of quanta is a dimensionless quantity, we should put $n_q = n_c$, and hence $n_q$ should be also infinitely large. This consideration suggests that if we extend quantum mechanics to the states with infinite number of quanta, the correspondence between quantum mechanics and classical mechanics becomes much more consistent and thorough. The rigorous mathematical description of this extension of quantum mechanics has been already possible by use of nonstandard analysis. In this section, the basic principles of nonstandard analysis is explained, and in the next section the macroscopic extension of quantum mechanics is described.

In nonstandard analysis, we consider a large collection $V$ of sets which contains every set used in the ordinary analysis. Note that $V$ contains a set of a set, a set of a set of a set, etc. Any set $M$ in $V$ has its nonstandard extension $^\star M$. A set $X$ which is an element of some $^\star M$ is called an *internal* set. We have two basic principles in nonstandard analysis, called the *transfer principle* and the *saturation principle*. The transfer principle states that any *elementary property* which holds for $M_1, M_2, \ldots, M_n$ also holds for $^\star M_1, {}^\star M_2, \ldots, {}^\star M_n$. A difficult point in nonstandard analysis is to understand the notion of an elementary property. The rigorous explanation of this notion needs a substantial part of mathematical logic. Fortunately, most of mathematical properties used in physics is elementary properties—a reason why the intuitive use of infinitesimal numbers has been useful in physics—and hence "any elementary property" can read "any property" in most of time. The saturation principle which we assume in this paper states that any sequence of internal sets with the finite intersection property has a nonempty intersection. For further detail on the basic framework of nonstandard analysis we shall refer to Hurd-Loeb [21]. Our framework is called an $\aleph_1$-saturated bounded elementary extension of a superstructure which contains $\mathbf{R}$ in the usual terminology.

Let $^\star \mathbf{N}$ be the nonstandard extension of the set $\mathbf{N}$ of natural numbers. Then $^\star \mathbf{N}$ satisfies all elementary properties of the natural number system by the transfer principle. An element of $^\star \mathbf{N}$ is called an *internal* natural number, and an element of $^\star \mathbf{N} \setminus \mathbf{N}$ is called a *nonstandard* natural number, where $^\star \mathbf{N} \setminus \mathbf{N}$ is the set of elements in $^\star \mathbf{N}$ which are not in $\mathbf{N}$. Every nonstandard natural number is greater than any (standard) natural number, and yet has all elementary properties of (standard) natural numbers. An internal set enumerated by a nonstandard natural number with an internal one-to-one function is called a *hyperfinite* set. Every hyperfinite set is an infinite set, but has all elementary properties of finite sets.

Let $^\star\mathbf{R}$ be the nonstandard extension of the set $\mathbf{R}$ of real numbers. Then, $^\star\mathbf{R}$ is a proper ordered field extension of $\mathbf{R}$ by the transfer principle. An element of $^\star\mathbf{R}$ is called a *hyperreal* number. The hyperreal number system $^\star\mathbf{R}$ is not complete nor satisfies the Archimedean axiom, since these properties are not "elementary properties". A hyperreal number $x$ is called *infinite* if $|x| > n$ for any $n \in \mathbf{N}$, *finite*, write $|x| < \infty$, if there is some $n \in \mathbf{N}$ such that $|x| < n$, and *infinitesimal* if $|x| < 1/n$ for any $n \in \mathbf{N}$. For any $x$, $y \in {}^\star\mathbf{R}$, we will write $x \approx y$ if $|x - y|$ is infinitesimal. For any finite hyperreal number $x$, there is a unique real number $r$ such that $r \approx x$; this $r$ is called the *standard part* of $x$ and denoted by $^\circ x$. Any function $f$ from $X$ to $Y$ is extended to an internal function $^\star f$ from $^\star X$ to $^\star Y$. A sequence $a_n \in \mathbf{R}$ ($n \in \mathbf{N}$) is extended to an internal sequence $^\star a_\nu \in {}^\star\mathbf{R}$ ($\nu \in {}^\star\mathbf{N}$), so that $\lim_{n \to \infty} a_n = a$ if and only if $^\star a_\nu \approx a$ for all $\nu \in {}^\star\mathbf{N} \setminus \mathbf{N}$. A function $f : \mathrm{dom}(f) \to \mathbf{R}$ with $\mathrm{dom}(f) \subset \mathbf{R}$ is continuous at $x \in \mathrm{dom}(f)$ if and only if $^\star f(x) \approx {}^\star f(y)$ for all $y \in \mathrm{dom}(f)$ with $x \approx y$. The above notions and results of infinitesimal analysis are naturally extended to the nonstandard extension $^\star\mathbf{C}$ of the complex number field $\mathbf{C}$.

Let $(E, p)$ be an internal normed linear space over $^\star\mathbf{C}$ with norm $p$. We define the *principal galaxy* $E_G$ of $(E, p)$ and the *principal monad* $E_M$ of $(E, p)$ as follows:

$$E_G \;=\; \{x \in E \mid p(x) < \infty\}, \tag{5.1}$$

$$E_M \;=\; \{x \in E \mid p(x) \approx 0\}. \tag{5.2}$$

Then both $E_G$ and $E_M$ are linear spaces over $\mathbf{C}$. Let $\hat{E} = E_G/E_M$ and $\hat{p}(\,^\circ x) = {}^\circ(p(x))$ for $x \in E_G$, where $^\circ x = x + E_M$. Then $(\hat{E}, \hat{p})$ becomes a normed linear space over $\mathbf{C}$, called the *nonstandard hull* of $(E, p)$. By the saturation principle, it is concluded that the nonstandard hull $(\hat{E}, \hat{p})$ is a Banach space [21, p. 156]. When $E$ has an inner product $\langle \cdot | \cdot \rangle$ which determines the norm $p$, the nonstandard hull $(\hat{E}, \hat{p})$ is a Hilbert space with the inner product such that $\langle\,{}^\circ x\,|\,{}^\circ y\rangle = {}^\circ\langle x|y\rangle$ for all $x$, $y \in E_G$.

## 6. Macroscopic extension of the Schrödinger representation

Let ${}^\star\mathcal{H}$ be the nonstandard extension of the Hilbert space $\mathcal{H}$ of quantum states, and $\widehat{{}^\star\mathcal{H}}$ its nonstandard hull. Then $\widehat{{}^\star\mathcal{H}}$ is a Hilbert space with inner product $\langle\,{}^\circ\xi\,|\,{}^\circ\eta\rangle = {}^\circ\langle\xi|\eta\rangle$ for $\xi$, $\eta \in {}^\star\mathcal{H}_G$. Let $\nu$ be a nonstandard natural number and $\mathcal{D}$ the internal $\nu$-dimensional subspace of ${}^\star\mathcal{H}$ spanned by the hyperfinite set $\{|n\rangle \mid n = 0, 1, \ldots, \nu - 1\}$. Then the nonstandard hull $\hat{\mathcal{D}}$ contains $\mathcal{H}$ as a closed subspace by the canonical isometric embedding $V_\mathcal{H}$ which maps $\xi \in \mathcal{H}$ to ${}^\circ{}^\star\xi \in \hat{\mathcal{D}}$, i.e., $\xi = V_\mathcal{H}\xi = {}^\circ{}^\star\xi$. The nonstandard hulls $\widehat{{}^\star\mathcal{H}}$ and $\hat{\mathcal{D}}$ are, thus, Hilbert spaces which satisfy the relations

$$\mathcal{H} \subset \hat{\mathcal{D}} \subset \widehat{{}^\star\mathcal{H}}. \tag{6.1}$$

Let $\mathcal{A}$ be the internal algebra of internal linear operators on $\mathcal{D}$. Then $\mathcal{A}$ is a hyperfinite dimensional internal *-algebra over ${}^\star\mathbf{C}$. For $x \in \mathcal{A}$, let $\|x\|$ be the internal uniform norm of $x$. Then, the nonstandard hull $\hat{\mathcal{A}}$ of $(\mathcal{A}, \|\cdot\|)$ becomes a Banach *-algebra with norm $\|{}^\circ x\| = {}^\circ\|x\|$. Then the norm $\|\cdot\|$ satisfies the C*-condition, i.e., $\|({}^\circ x)^* ({}^\circ x)\| = \|{}^\circ x\|^2$ for all ${}^\circ x \in \hat{\mathcal{A}}$, and hence $(\hat{\mathcal{A}}, \|\cdot\|)$ is a C*-algebra.

Any internal operator $x \in \mathcal{A}_G$ leaves $\mathcal{D}_G$ and $\mathcal{D}_M$ invariant and gives rise to a bounded operator $\pi(x)$ on $\hat{\mathcal{D}}$ such that $\pi(x)\,{}^\circ\xi = {}^\circ(x\xi)$ for all $\xi \in \mathcal{D}_G$. Then the correspondence ${}^\circ x \mapsto \pi(x)$ for $x \in \mathcal{A}_G$ defines a faithful *-representation of the C*-algebra $\hat{\mathcal{A}}$ on the Hilbert space $\hat{\mathcal{D}}$.

Let $P_\mathcal{D}$ be the internal projection from ${}^\star\mathcal{H}$ onto $\mathcal{D}$. Any bounded operator $T$ on $\mathcal{H}$ has the nonstandard extension ${}^\star T$ which is an internal bounded linear operator on ${}^\star\mathcal{H}$. Denote the restriction of $P_\mathcal{D}\,{}^\star T$ to $\mathcal{D}$ by ${}^\star T_\mathcal{D}$. Then it is easy to see that ${}^\star T_\mathcal{D} \in \mathcal{A}_G$. We denote the operator $\pi({}^\star T_\mathcal{D})$ on $\hat{\mathcal{D}}$ by $T_\mathcal{D}$, which is called the *standard hyperfinite extension* of $T$ to $\hat{\mathcal{D}}$.

Then $T_{\mathcal{D}} = T$ on $\mathcal{H}$ and $\|T_{\mathcal{D}}\| = \|T\|$. Properties of such extensions from $\mathcal{H}$ to $\hat{\mathcal{D}}$ are studied by Moore [38] extensively. Now the following statement is easily established; cf. [38, Lemma 1.3].

**Theorem 6.1.** *The mapping $\mathcal{E}_{\mathcal{D}} : T \mapsto T_{\mathcal{D}}$ is a completely positive isometric injection from $\mathcal{L}(\mathcal{H})$ to $\hat{\mathcal{A}}$ and the mapping $\mathcal{E}_{\mathcal{H}} : \pi(T) \mapsto V_{\mathcal{H}}^{\dagger} \pi(T) V_{\mathcal{H}}$ is a completely positive surjection from $\hat{\mathcal{A}}$ onto $\mathcal{L}(\mathcal{H})$. The composition $\mathcal{E}_{\mathcal{H}} \mathcal{E}_{\mathcal{D}}$ is the identity map on $\mathcal{L}(\mathcal{H})$, and the composition $\mathcal{E}_{\mathcal{D}} \mathcal{E}_{\mathcal{H}}$ is the norm one projection from $\hat{\mathcal{A}}$ onto $\mathcal{E}_{\mathcal{D}}(\mathcal{L}(\mathcal{H}))$.*

Let $A$ be an observable represented by a bounded Hermitian operator $\hat{A}$ on the Hilbert space $\mathcal{H}$ of the Schrödinger representation. Then as described by Theorem 6.1, the operator $\hat{A}$ is extended to the bounded Hermitian operator $\hat{A}_{\mathcal{D}}$ on the space $\hat{\mathcal{D}}$. Thus, we have a new quantization which associates an observable $A$ with a Hermitian operator on $\hat{\mathcal{D}}$. The above theorem states that this quantization leaves the mean values of observables in quantum states of the Schrödinger representation unchanged, i.e.,

$$\langle \psi | \hat{A} | \psi \rangle = \langle \psi | \hat{A}_{\mathcal{D}} | \psi \rangle,$$

for any bounded observable $A$ and state $\psi \in \mathcal{H}$.

Our claim is that there is a physical quantity such that its Hermitian operator on $\mathcal{H}$ does not exists, but its Hermitian operator on $\hat{\mathcal{D}}$ does exist. In the next section, we show that the phase is one of such physical quantities. In Section 8, we show the macroscopic character of states in $\hat{\mathcal{D}}$ orthogonal to $\mathcal{H}$, which suggests the semiclassical nature of the observables which fail to have the Hermitian operator on $\mathcal{H}$.

## 7. The Hermitian phase operator

Let $\Delta\theta = 2\pi/\nu$, and $\theta_m = m\Delta\theta$ for each $m$ $(m = 0, 1, \ldots, \nu - 1)$. The *internal phase eigenstate* $|\theta_m\rangle$ in $\mathcal{D}$ is defined by

$$|\theta_m\rangle = \nu^{-1/2} \sum_{n=0}^{\nu-1} e^{in\theta_m} |n\rangle. \tag{7.1}$$

Then we have

$$\langle\theta_m|\theta_{m'}\rangle = \delta_{m,m'}. \tag{7.2}$$

The *internal phase operator* $\hat{\phi}_I$ on $\mathcal{D}$ is defined by

$$\hat{\phi}_I = \sum_{m=0}^{\nu-1} \theta_m |\theta_m\rangle\langle\theta_m|. \tag{7.3}$$

Then the internal phase operator $\hat{\phi}_I$ has the internal spectrum $\{2m\pi/\nu \mid m = 0, 1, \ldots, \nu-1\}$ and hence is in $\mathcal{A}_G$. Thus we have the Hermitian operator $\pi(\hat{\phi}_I)$ on $\hat{\mathcal{D}}$, denoted by $\hat{\phi}$ and called the *Hermitian phase operator* on $\hat{\mathcal{D}}$. Denote by $\Lambda(\hat{\phi})$ the spectrum of $\hat{\phi}$ and $\Pi_0(\hat{\phi})$ the point spectrum (eigenvalues) of $\hat{\phi}$.

**Theorem 7.1.** *We have $\Lambda(\hat{\phi}) = \Pi_0(\hat{\phi}) = [0, 2\pi]$. For each $\theta \in \mathbf{R}$ $(0 \leq \theta \leq 2\pi)$, the vector $^\circ|\theta_m\rangle \in \hat{\mathcal{D}}$ with $\theta_m \approx \theta$ is an eigenvector of $\hat{\phi}$ for the eigenvalue $\theta$.*

For each $\theta \in \mathbf{R}$ $(0 \leq \theta \leq 2\pi)$ and $n \in \mathbf{N}$, define $F(\theta, n)$ to be the internal projection

$$F(\theta, n) = \sum_{\theta_m \leq \theta + n^{-1}} |\theta_m\rangle\langle\theta_m|.$$

Then $F(\theta, n) = 0$ if $\theta + n^{-1} < 0$, and for each $\theta$ the sequence $\pi(F(\theta, n))$ $(n \in \mathbf{N})$ is a monotone decreasing sequence of projections on $\hat{\mathcal{D}}$. Define $E_\phi(\theta)$ to be the strong limit of $\pi(F(\theta, n))$. Then $E_\phi(\theta)$ $(\theta \in [0, 2\pi])$ is the spectral resolution for $\hat{\phi}$ [38, Theorem 4.1]. Now it is shown in the next theorem that the Hermitian phase operator $\hat{\phi}$ has the desirable properties.

**Theorem 7.2.** *The Hermitian phase operator $\hat{\phi}$ on $\hat{\mathcal{D}}$ with its spectral resolution $E_\phi$ satisfies the following conditions* (P1)–(P4)*:*

(P1) *The spectral resolution $E_\phi$ of $\hat{\phi}$ is a Naimark extension of the phase POM $P_\phi$, i.e.,*

$$\langle\psi|E_\phi(\theta)|\psi\rangle = \langle\psi|P_\phi(\theta)|\psi\rangle,$$

*for any state $\psi \in \mathcal{H}$.*

(P2) *For any continuous function $f$, the Hermitian operator $f(\hat{\phi})$ is the limit of $f(\hat{\phi}_s)$ as $s$ tends to infinity in the weak operator topology of $\mathcal{H}$, where $\hat{\phi}_s$ is the Pegg-Barnett operator, i.e.,*

$$\lim_{s\to\infty}\langle\psi|f(\hat{\phi}_s)|\psi\rangle = \langle\psi|f(\hat{\phi})|\psi\rangle,$$

*for all states $\psi \in \mathcal{H}$.*

(P3) *The Susskind-Glogower phase operators are given by the relations:*

$$\begin{aligned}
\widehat{\exp}_{SG}\pm i\phi &= V_{\mathcal{H}}^\dagger \exp\pm i\hat{\phi}\, V_{\mathcal{H}},\\
\widehat{\cos}_{SG}\phi &= V_{\mathcal{H}}^\dagger \cos\hat{\phi}\, V_{\mathcal{H}},\\
\widehat{\sin}_{SG}\phi &= V_{\mathcal{H}}^\dagger \sin\hat{\phi}\, V_{\mathcal{H}}.
\end{aligned}$$

(P4) *For any continuous function $f$,*

$$\begin{aligned}
\langle{}^\circ\psi|f(\hat{\phi})|{}^\circ\psi\rangle &\approx \langle\psi|{}^\star f(\hat{\phi}_I)|\psi\rangle\\
&= \sum_{m=0}^{\nu-1} {}^\star f(\theta_m)|\langle\theta_m|\psi\rangle|^2,
\end{aligned}$$

*for all $\psi \in \mathcal{D}_G$.*

As consequences from (P1), we obtain the statistics of the phase only from the ordinary quantum rules with the Hermitian operator $\hat{\phi}$. Let $\psi \in \mathcal{H}$ be a normalized state of the system.

Then $\psi$ is in the domain of $\hat{\phi}$ by relation (6.1). The probability distribution function of the phase in the state $\psi$ is given by

$$\Pr[\phi \leq \theta | \psi] = \langle \psi | E_\phi(\theta) | \psi \rangle.$$

For any Borel function $f$, the physical quantity $f(\phi)$ has the Hermitian operator

$$f(\hat{\phi}) = \int_0^{2\pi} f(\theta) \, dE_\phi(\theta),$$

by the function calculus, and the mean value of $f(\phi)$ in the state $\psi$ is given by

$$\mathrm{Ex}[f(\phi) | \psi] = \langle \psi | f(\hat{\phi}) | \psi \rangle.$$

## 8. Macroscopic states

We have extended the Schrödinger representation on $\mathcal{H}$ to the hyperfinite dimensional space $\hat{\mathcal{D}}$. The following theorem shows that the states in $\hat{\mathcal{D}} \ominus \mathcal{H}$ can be interpreted naturally as the classical limits of the quantum mechanical states in $\mathcal{H}$.

**Theorem 8.1.** *Let $T \in \mathcal{L}(\mathcal{H})$. Suppose that $\langle n|T|n'\rangle$ $(n, \ n' \in \mathbf{N})$ is a Cauchy sequence in $n$ and $n'$. Then for any nonstandard $k, \ k' \in {}^\star\mathbf{N}$ with $|k\rangle, \ |k'\rangle \in \mathcal{D}$ the standard hyperfinite extension $T_\mathcal{D} \in \hat{\mathcal{A}}$ of $T$ satisfies the relation*

$$\langle k|T_\mathcal{D}|k'\rangle = \lim_{n,n'\to\infty} \langle n|T|n'\rangle. \tag{8.1}$$

**Remark.** Suppose that the nonstandard universe is constructed by a bounded ultrapower of a superstructure based on $\mathbf{R}$, with the index set $I = \mathbf{N}$ and a free ultrafilter $\mathcal{U}$. Then any nonstandard number $k \in {}^\star\mathbf{N} \setminus \mathbf{N}$ is represented by a sequence $s(i)$ $(i \in \mathbf{N})$ of natural numbers in such a way that two sequences $s(i)$ and $s'(i)$ represents the same nonstandard number $k$ if and only if $\{i \in \mathbf{N} | \ s(i) = s'(i)\} \in \mathcal{U}$. Let $m$ be a nonstandard number corresponding to a sequence $s(i)$. Let $a(n)$ be a bounded sequence of complex numbers.

Then the standard part of $^\star a(m)$ coincides with the ultralimit of the subsequence $a(s(i))$ of $a(n)$, i.e.,

$$^\circ(^\star a(m)) = \lim_{i \to \mathcal{U}} a(s(i)). \tag{8.2}$$

Thus, for any $T \in \mathcal{L}(\mathcal{H})$ and $k,\ k' \in {}^\star\mathbf{N} \setminus \mathbf{N}$, we have,

$$\langle k | T_{\mathcal{D}} | k' \rangle = \lim_{i,j \to \mathcal{U}} \langle s(i) | T | s'(j) \rangle, \tag{8.3}$$

provided $k,\ k'$ are represented by sequences $s(i),\ s'(j)$. Thus the matrix element $\langle k | T_{\mathcal{D}} | k' \rangle$ is the ultralimit of a subsequence of $\langle n | T | n' \rangle$ $(n,\ n' \in \mathbf{N})$, even if it is not a Cauchy sequence.

Let $A$ be an observable represented by a bounded Hermitian operator $\hat{A}$ on the Hilbert space $\mathcal{H}$ of the Schrödinger representation. Since $\hat{A}_{\mathcal{D}}$ is an extension, the matrix element $\langle n | \hat{A}_{\mathcal{D}} | n' \rangle$ for standard $n,\ n' \in \mathbf{N}$ is the same as that of the operator $\hat{A}$, i.e., $\langle n | \hat{A}_{\mathcal{D}} | n' \rangle = \langle n | \hat{A} | n' \rangle$. From Theorem 8.1, the matrix element $\langle n | \hat{A}_{\mathcal{D}} | n' \rangle$ for nonstandard $n,\ n'$ has also a clear physical interpretation, i.e., it is the classical limit of the matrix elements $\langle n | \hat{A} | n' \rangle$. Thus, the states in $\mathcal{H} \setminus \hat{\mathcal{D}}$ are naturally considered as the classical limits of quantum states. We shall call $\mathcal{H}$ the *purely microscopic* part of $\hat{\mathcal{D}}$, and $\hat{\mathcal{D}} \ominus \mathcal{H}$ the *purely macroscopic* part of $\hat{\mathcal{D}}$. The space $\hat{\mathcal{D}}$ contains microscopic states in $\mathcal{H}$, macroscopic states in $\hat{\mathcal{H}} \ominus \hat{\mathcal{D}}$, and superpositions of those states in $\hat{\mathcal{D}}$. Thus, it is appropriate to call the extension $\mathcal{H} \to \hat{\mathcal{D}}$ with the extensions $\hat{A} \to \hat{A}_{\mathcal{D}}$ for observables $A$ as the *macroscopic extension* of the Schrödinger representation. The greatest significance is, of course, that we can associate the phase with a Hermitian operator on the macroscopic extension.

## 9. Conclusions

The claim, "We have recently shown that a Hermitian optical phase operator exists. This result contradicts the well established belief that no such operator can be constructed." due to Barnett and Pegg [5] is not appropriate. Instead, they construct a sequence of Hermitian operators $\hat{\phi}_s$ on finite $s$-dimensional spaces such that the statistics obtained by the operators $\hat{\phi}_s$ approaches to the statistics of the phase as $s$ tends to infinity. Their statistics of the phase obtained by this limit process is consistent with the one obtained from the phase POM.

In this paper, we give a demonstration of the claim that the correct statistics of the phase is the one given by the phase POM as well as the Pegg-Barnett limit process, and construct a single Hermitian operator $\hat{\phi}$ on an infinite dimensional Hilbert space which gives the statistics of the phase by the ordinary quantum rules. Furthermore, the Hermitian operator $\hat{\phi}$ is infinitesimally close to the internal phase operator $\hat{\phi}_I$, that is, the statistical prediction given by $\hat{\phi}$ is obtained from $\hat{\phi}_I$ with only infinitesimal difference. The remarkable character of the internal phase operator $\hat{\phi}_I$ is that this operator shares all elementary properties of the Pegg-Barnett operators on finite dimensional spaces. This follows from the transfer principle in nonstandard analysis.

Thus the statistics of the phase variable is obtained by the following procedure. Fix a nonstandard natural number $\nu$ and consider the $\nu$-dimensional internal linear space $\mathcal{D}$. The internal phase operator $\hat{\phi}_I$ is well-defined on $\mathcal{D}$. Recall that all number states and all phase states are contained in $\mathcal{D}$ (cf. Theorem 7.1). Every computation involving $\hat{\phi}_I$ is formally the same as the computation involving the Pegg-Barnet operator, but gives the correct statistics with only infinitesimal error. The last step is, if desire, to remove infinitesimals from the result of the computation.

Theorem 7.2 states that this procedure is equivalent to the computation with the usual quantum rules involving the Hermitian phase operator $\hat{\phi}$ on the Hilbert space $\hat{\mathcal{D}}$, which extends the Hilbert space $\mathcal{H}$ of quantum states by adding macroscopic states. This result does not contradict the well established belief that no such operator can be constructed on the Hilbert space of *quantum states*. Instead, we have shown that the phase operator can be constructed on the Hilbert space of the quantum states plus the *macroscopic states*.

**Acknowledgments**

I thank Professor Horace P. Yuen for his warm hospitality at Northwestern University where the final version of the manuscript is prepared.

## A. Appendices

### A.1. Proof of Theorem 7.1

Let $\theta \in [0, 2\pi]$ and $\theta_m \approx \theta$. Then obviously,

$$\hat{\phi}\,{}^\circ|\theta_m\rangle = {}^\circ(\hat{\phi}_I|\theta_m\rangle) = {}^\circ(\theta_m|\theta_m\rangle) = {}^\circ\theta_m\,{}^\circ|\theta_m\rangle = \theta\,{}^\circ|\theta_m\rangle,$$

and hence ${}^\circ|\theta_m\rangle$ is an eigenvector of $\hat{\phi}$ corresponding to eigenvalue $\theta$. Thus we have $[0, 2\pi] \subset \Pi_0(\hat{\phi})$. Since $0 \leq \theta_m \leq 2\pi$, we have $0 \leq \langle\psi|\hat{\phi}_I|\psi\rangle \leq 2\pi$ for any unit vector $\psi \in \mathcal{D}$, and hence

$$0 \leq \langle\,{}^\circ\psi|\hat{\phi}|\,{}^\circ\psi\rangle = {}^\circ\langle\psi|\hat{\phi}_I|\psi\rangle \leq 2\pi.$$

It follows that $0 \leq \hat{\phi} \leq 2\pi 1$, so that $\Lambda(\hat{\phi}) \subset [0, 2\pi]$. Therefore, $\Lambda(\hat{\phi}) = [0, 2\pi] = \Pi_0(\hat{\phi})$. This completes the proof.

### A.2. Proof of Theorem 7.2

Let $n, n' \in \mathbf{N}$. We have

$$
\begin{aligned}
\langle n|E_\phi(\theta)|n'\rangle &= \lim_{k\to\infty} \langle n|\hat{F}(\theta, k)|n'\rangle \\
&= \lim_{k\to\infty} {}^\circ\langle n|F(\theta, k)|n'\rangle \\
&= \lim_{k\to\infty} {}^\circ \sum_{\theta_m \leq \theta + k^{-1}} e^{i(n-n')\theta_m} \frac{\Delta\theta}{2\pi} \\
&= \lim_{k\to\infty} \int_0^{\theta+k^{-1}} e^{i(n-n')\bar{\theta}} \frac{d\bar{\theta}}{2\pi} \\
&= \int_0^\theta e^{i(n-n')\bar{\theta}} \frac{d\bar{\theta}}{2\pi}.
\end{aligned}
$$

From (3.12), this concludes condition (P1). Let $a_s$ ($s \in \mathbf{N}$) be a sequence such that $a_s = \langle n|f(\hat{\phi}_s)|n'\rangle$, and ${}^\star a_s$ ($s \in {}^\star\mathbf{N}$) its nonstandard extension. Then we have $\lim_{s\to\infty} a_s \approx {}^\star a_\nu$, and hence

$$
\begin{aligned}
\lim_{s\to\infty} \langle n|f(\hat{\phi}_s)|n'\rangle \;\; &\approx \;\; \langle n|{}^\star f(\hat{\phi}_I)|n'\rangle \\
&= \;\; \sum_{m=0}^{\nu-1} {}^\star f(\theta_m) e^{i(n-n')\theta_m} \frac{\Delta\theta}{2\pi} \\
&\approx \;\; \int_0^{2\pi} f(\theta) e^{i(n-n')\theta} \frac{d\theta}{2\pi} \\
&= \;\; \langle n|f(\hat{\phi})|n'\rangle.
\end{aligned}
$$

Since the first and the last term is standard, they must be the same, i.e.,

$$
\lim_{s\to\infty} \langle n|f(\hat{\phi}_s)|n'\rangle = \langle n|f(\hat{\phi})|n'\rangle.
$$

Then (P2) follows easily, To prove (P3), by the obvious relations, it suffices to show the first equation. We have

$$
\begin{aligned}
\langle n|e^{-i\hat{\phi}}|n'\rangle \;\; &= \;\; \int_0^{2\pi} e^{-i\theta} \langle n|dE_\phi(\theta)|n'\rangle \\
&= \;\; \int_0^{2\pi} e^{i(n-n'-1)\theta} \frac{d\theta}{2\pi} \\
&= \;\; \delta_{n,n'+1},
\end{aligned}
$$

whence the desired equation is obtained from the relation

$$
\begin{aligned}
\langle n|(\hat{N}+1)^{-\frac{1}{2}}\hat{a}|n'\rangle \;\; &= \;\; \langle n|n'+1\rangle \\
&= \;\; \delta_{n,n'+1}.
\end{aligned}
$$

Thus (P3) follows easily. (P4) is obvious, and the proof is completed.

### A.3. Proof of Theorem 8.1

Let $a_{n,n'} = \langle n|T|n'\rangle$ for $n$, $n' \in \mathbf{N}$, and $L = \lim_{n,n'\to\infty} a_{n,n'}$. Let ${}^\star a_{m,m'}$ ($m$, $m' \in {}^\star\mathbf{N}$) be the nonstandard extension of the sequence $a_{n,n'}$ ($n$, $n' \in \mathbf{N}$). Then, $L \approx a_{m,m'}$ for all $m,\ m' \in {}^\star\mathbf{N} \setminus \mathbf{N}$. Let $k,\ k' \in \{0, 1, \ldots, \nu - 1\} \setminus {}^\star\mathbf{N}$. By transfer principle, ${}^\star a_{k,k'} = \langle k| {}^\star T |k'\rangle$, and hence $L \approx \langle k| {}^\star T |k'\rangle$. It follows that $L = {}^\circ \langle k| {}^\star T |k'\rangle = \langle k|T_D|k'\rangle$. This completes the proof.